\def\cA{{\cal A}}

\def\plushc{\ + h.c.}

\newcommand{\I}{{\rm i}}

\newcommand{\cL}{{\cal L}}
\newcommand{\cD}{{\cal D}}
\newcommand{\cF}{{\cal F}}

\newcommand{\nn}{\nonumber}
\newcommand{\al}[1]{\begin{eqnarray}#1\end{eqnarray}}
\newcommand{\eq}[1]{\begin{equation}#1\end{equation}}

\def\be{\begin{equation}}
\def\ee{\end{equation}}
\def\bea{\begin{eqnarray}}
\def\eea{\end{eqnarray}}

\def\cropen#1{\crcr\noalign{\vskip #1}}
\def\crr{\cropen{1\jot }}
\def\Red{}
\def\Black{}
\documentstyle[epsf, 12pt]{article}
\begin{document}

\thispagestyle{empty}
\begin{flushright}
JHU-TIPAC-99005\\
hep-th/9904213\\
\end{flushright}

\bigskip\bigskip\begin{center} {\bf
\LARGE{\bf Dual Supersymmetry Algebras from Partial Supersymmetry
Breaking} }
\end{center} \vskip 1.0truecm

\centerline{\bf Richard Altendorfer and Jonathan Bagger}

\vskip5mm
\centerline{\it Department of Physics and Astronomy}
\centerline{\it The Johns Hopkins University}
\centerline{\it 3400 North Charles Street}
\centerline{\it Baltimore, MD 21218, U.S.A.}
\vskip5mm

\bigskip \nopagebreak \begin{abstract}
\noindent The partial breaking of supersymmetry in flat space can
be accomplished using any one of three dual representations for
the massive $N=1$ spin-3/2 multiplet.  Each of the representations
can be ``unHiggsed'', which gives rise to a set of dual $N=2$
supergravities and supersymmetry algebras. 
\end{abstract}

\newpage\setcounter{page}1

\section{Introduction}

The transition from string theory to the standard model can
be characterized in terms of a hierarchy of supersymmetries:
from $N=8$ to $N=0$.  These supersymmetries must be
spontaneously broken, either all at once, to $N=0$, or
partially, first to $N=1$ (or higher) and then to $N=0$. 

For phenomenological applications of weak-scale supersymmetry,
one would like to construct the effective field theory that
describes the breaking of $N=8$ to $N=1$.  In this paper we
will focus on a simpler case, that of $N=2$ broken to $N=1$.  
We will construct a set of effective supergravity theories
that contain an  unbroken, linearly realized $N=1$ supersymmetry,
as well as a spontaneously broken, nonlinearly realized, $N=2$.

Heuristically, it might seem impossible to partially break
$N=2$ to $N=1$.  The argument runs as follows.  Start with
the $N=2$ supersymmetry algebra
\bea
\{ Q_\alpha,\,\bar
Q_{\dot\alpha} \} &\ =\ & 2\, \sigma^m_{
\alpha\dot\alpha}\,P_m \nonumber \\
\{ S_\alpha,\,\bar
S_{\dot\alpha} \} &=& 2\, \sigma^m_{
\alpha\dot\alpha}\,P_m\ ,
\eea
where $Q_\alpha$ and its conjugate $\bar Q_{\dot\alpha}$ denote the
first, unbroken supersymmetry, and $S_\alpha$, $\bar S_{\dot\alpha}$
the second.  Suppose that one supersymmetry is not broken, so
\be
Q\, |0\rangle \ =\ \bar Q\, |0\rangle\ =\ 0\ .
\ee
Because of the supersymmetry algebra, this implies that the Hamiltonian
also annihilates the vacuum,
\be
H\, |0\rangle \ =\ 0\ .
\ee
Then, according to the supersymmetry algebra,
\be
(\bar S S + S \bar S)\, |0\rangle\ =\ 0\ .
\label{SSbar}
\ee
For a positive definite Hilbert space, this leads one to
conclude
that
\be
S\, |0\rangle\ =\ \bar S\, |0\rangle\ =\ 0\ .
\ee

This argument can be evaded by two loopholes.  
The first is that in a spontaneously broken theory, one can only
consider the algebra of the {\it currents}, since the charges of
the spontaneously broken symmetries do not exist rigorously. 
The second exploits the fact that in covariantly-quantized
supergravity theories, the gravitino $\psi_{m\alpha}$ is a gauge field
with negative-norm components, so the Hilbert space does not have
positive norm.

There are by now many examples of partial supersymmetry breaking
which take advantage the first loophole.  The first was given by
Hughes, Liu and Polchinski \cite{hlp}, who showed that
supersymmetry is partially broken on the world volume of an
$N=1$ supersymmetric 3-brane propagating in six-dimensional 
superspace.  Later, Bagger and Galperin \cite{2bagger,BGT}
used the techniques of Coleman, Wess, Zumino \cite{cwz}, and 
Volkov \cite{volkov} to construct an effective field theory
of partial supersymmetry breaking, with the broken supersymmetry
realized nonlinearly.  They found that the Goldstone fermion
could belong to an $N=1$ chiral {\it or} an $N=1$ vector multiplet. 
Antoniadis, Partouche and Taylor discovered another realization
in which the Goldstone fermion is contained in an $N=2$ vector
multiplet \cite{apt}.

Each of these examples relies on the fact that in partially broken
supersymmetry, the current algebra can be modified as follows,
\bea
\{ \bar Q_{\dot\alpha},\
J^1_{\alpha m}
\} &\ =\ & 2 \,\sigma^n_{\alpha\dot\alpha}\,
T_{mn}\nonumber \\
\{ \bar S_{\dot\alpha},\ J^2_{\alpha m}
\} &=& 2\, \sigma^n_{\alpha\dot\alpha}\,
(v^4 \eta_{mn}+ T_{mn})\ ,\label{zweite}
\eea
where the $J^i_{\alpha m}$ ($i = 1,2$) are the supercurrents and
$T_{mn}$ is the stress-energy tensor.  The shift in the second
stress-energy tensor in (\ref{zweite}) prevents the current algebra
from being integrated into a charge algebra, and circumvents the
no-go theorem.

In gravity, however, a shift in the stress-energy tensor corresponds
to a shift in the vacuum energy.  This suggests that the mechanism
of partial breaking might be different in supergravity theories.
Indeed, theories with partial breaking were constructed by Cecotti,
Girardello and Porrati, and by Zinov'ev \cite{zin}, starting from
linearly realized $N=2$ supergravity.  (A geometrical interpretation
was given in \cite{fgp-local}.)  These authors considered scenarios
with vector- and hypermultiplets and found that the gravitational
couplings exploited the second loophole.  It is natural to ask
whether their results apply more generally in supergravity
theories.

In this paper we will address this question using a model-independent
approach with a minimal field content motivated by the superHiggs-effect.
We will see that partial breaking in flat space can be accomplished
using three dual representations for the $N=1$ massive spin-3/2
multiplet.  When coupled to gravity, the dual representations
give rise to new $N=2$ supergravities with new $N=2$ supersymmetry
algebras.

In each case, our technique will be as follows:  We will start with
the Lagrangian and supersymmetry transformations for the massive
$N=1$ spin-3/2 multiplet.  We shall then ``unHiggs" the
representation by adding appropriate Goldstone fields and coupling
it to gravity.

\section{The SuperHiggs Effect in Partially Broken Supersymmetry}

\subsection{Dual Versions of Massive $N=1$ Spin-3/2 Multiplets}

The starting point for our investigation is the massive $N=1$
spin-3/2 multiplet. This multiplet contains six bosonic and
six fermionic degrees of freedom, arranged in states of the
following spins,
\be
\pmatrix{
{3\over2} \crr
1\ \ \ 1 \crr
{1\over2}}\ .
\ee
The traditional representation of this multiplet contains
the following fields \cite{fvn}:  one spin-3/2 fermion, one spin-1/2
fermion, and two spin-one vectors, each of mass $m$.  The dual
representations have the same fermions, but one or two antisymmetric
tensors in place of one or two of the vectors.  As we shall see,
each representation gives rise to a distinct $N=2$ supersymmetry
algebra.

The traditional representation is described by the following
Lagrangian \cite{fvn},
\bea
\cL &\ = \ & \epsilon^{m n \rho \sigma} \overline \psi_{m}
  \overline \sigma_n \partial_\rho \psi_\sigma 
 - \I \overline \zeta \overline \sigma^m \partial_m \zeta
 - {1 \over 4} \cA_{m n} \bar\cA^{m n} \nonumber \\
& &-\ {1\over 2}m^2\, \cA_m \bar\cA^m  
\ +\ {1\over 2}m\,\zeta\zeta  \ +\ {1\over 2}m\,\bar\zeta\bar\zeta \nonumber \\[1mm]
& & -\ m\,\psi_m \sigma^{m n} \psi_n -\ m\,\bar\psi_m \bar\sigma^{m n} \bar\psi_n\ .
\eea
Here $\psi_m$ is a spin-3/2 Rarita-Schwinger field, $\zeta$ a spin-1/2
fermion, and $\cA_m = A_m + \I B_m$ a complex spin-one vector.  This
Lagrangian is invariant under the following $N=1$ supersymmetry
transformations,
\bea
\delta_\eta \cA_m &\ =\ & 2\psi_m\eta - \I{2\over\sqrt{3}}
\bar\zeta\bar\sigma_m\eta 
-{2\over \sqrt{3}m}\partial_m(\zeta\eta) \nonumber \\
\delta_\eta \zeta &=& {1\over\sqrt{3}}\bar\cA_{mn}\sigma^{mn}
\eta -\I{m\over\sqrt{3}}
\sigma^m\bar\eta \cA_m \nonumber \\
\delta_\eta \psi_m &=& {1\over 3m}\partial_m(\bar\cA_{rs}
\sigma^{rs} \eta + 2\I m 
\sigma^n\bar\eta \cA_n)  
- {\I\over 2}(H_{+mn}\sigma^n + {1\over 3}H_{-mn}\sigma^n)
\bar\eta \nonumber
\\ & & -\ {2\over 3}m({\sigma_m}^n \bar\cA_n \eta + \bar
\cA_m\eta)\ ,
\eea
where $H_{\pm mn}=\cA_{mn}\pm {\I\over 2}\epsilon_{mnrs}\cA^{rs}$ and 
$\cA_{mn}=\partial_m\cA_n - \partial_n\cA_m$.

A dual Lagrangian and its supersymmetry transformations can
be found by using a Poincar\'e duality which relates a massive vector field
to a massive antisymmetric tensor field of rank two.  This duality
can be used to relate the vector $B_m$ to an antisymmetric tensor 
$B_{mn}$ by $B_{mn} = 1/m\ \epsilon_{mnrs}\partial^r B^{s}$
or $B_m =  v_m/m$ \cite{dual}.

This dual representation is special in the sense that it can
also be written in $N=1$ superspace formulation\footnote{
The massive version of the de Wit/van Holten formulation
(see \cite{gatsie} and references therein) leads to a
reducible supersymmetry representation.} \cite{ogsok}. 
It has the following component Lagrangian,
\bea
\cL &\ =\ & \epsilon^{pqrs} \bar \psi_{p}
  \bar \sigma_q \partial_r \psi_s
 - \I \bar \zeta \bar \sigma^m \partial_m \zeta
 - {1 \over 4} A_{mn} A^{mn} 
 + {1\over 2}v^m v_m \nonumber \\
& & -\ {1\over 2}m^2 A_m A^m  - {1\over4}m^2
B_{mn}B^{mn} \ +\ {1\over 2}m\,\zeta\zeta  \ +\ {1\over
2}m\,\bar\zeta\bar\zeta\nonumber\\[1mm]
& & -\ m\,\psi_m \sigma^{mn} \psi_n \ -\ m\,\bar\psi_m 
\bar\sigma^{mn} \bar\psi_n\ ,
\eea
where $A_{mn}$ is the field strength associated with the real vector
field $A_m$, and $v_m = {1\over 2}\epsilon_{mnrs} \partial^n B^{rs}$ 
is the field strength for the antisymmetric tensor
$B_{mn}$.  This Lagrangian is invariant under the following
$N=1$ supersymmetry transformations:\footnote{Here, 
the square brackets denote antisymmetrization, without a
factor of 1/2.}
\bea
\delta_\eta A_m &\ =\ & (\psi_m\eta + \bar\psi_m\bar\eta) + 
{\I\over\sqrt{3}}
(\bar\eta\bar\sigma_m\zeta - \bar\zeta\bar\sigma_m\eta)
-{1\over \sqrt{3}m}\partial_m(\zeta\eta +  \bar\zeta\bar\eta) 
\nonumber \\
\delta_\eta B_{mn} &=& {2\over\sqrt{3}}\left( \eta\sigma_{mn}\zeta 
+ {\I\over 2m}\partial_{[m}\bar\zeta\bar\sigma_{n]}\eta \right) 
\ +\ \I\eta\sigma_{[ m}\bar\psi_{n ]} + {1\over
m} \eta\psi_{mn}
\plushc \nonumber\\
\delta_\eta \zeta &=& {1\over\sqrt{3}} A_{mn}\sigma^{mn}\eta - 
{\I m\over\sqrt{3}}
 \sigma^m\bar\eta A_m - {1\over\sqrt{3}}m\sigma_{mn}\eta B^{mn} - 
{1\over\sqrt{3}} v_m\sigma^m\bar\eta \nonumber \\
\delta_\eta \psi_m &=& {1\over 3m}\partial_m \left( A_{rs}
\sigma^{rs}\eta + 2\I m
\sigma^n\bar\eta A_n  \right) 
 -{\I\over 2} (H^A_{+mn}\sigma^n + {1\over 3}H^A_{-mn}\sigma^n)
\bar\eta \nonumber \\
& & -\ {2\over 3}m ({\sigma_m}^n A_n \eta +
A_m\eta)\ +\ {1\over 3m}\partial_m \left( 2
v_n\sigma^n\bar\eta  - m 
 \sigma^{rs}\eta B_{rs} \right) \nonumber\\
&&    -\ {2\I\over 3} (v_m + \sigma_{mn}v^n)\eta 
 - {\I m\over 3} (B_{mn}\sigma^n\bar\eta + \I
\epsilon_{mnrs}B^{n r}\sigma^s\bar\eta) \ ,
\eea

A third representation can be found by dualizing the remaining
vector, $A_n$.  Its derivation is straightforward, so we will not
write its Lagrangian and transformations here.

Each of the three dual Lagrangians describe the dynamics of free
massive spin-3/2 and 1/2 fermions, together with their supersymmetric
partners, massive spin-one vector and tensor fields.  They can be
regarded as ``unitary gauge'' representations of theories with
additional symmetries:  a fermionic gauge symmetry for the massive
spin-3/2 fermion, as well as additional gauge symmetries associated
with the massive gauge fields.

\subsection{UnHiggsing Massive $N=1$ Spin-3/2 Multiplets}

To study partial breaking, these Lagrangians must be unHiggsed
by including appropriate gauge and Goldstone fields.  In each
case we need to add a Goldstone fermion and Goldstone bosons 
and then gauge the full $N=2$ supersymmetry.  In this way we can
construct theories with $N=2$ supersymmetry nonlinearly realized,
and $N=1$ represented linearly on the fields. The resulting effective
field theories describe the physics of partial supersymmetry breaking,
well below the scale $v$ where the second supersymmetry is broken.

In what follows we will focus on the first two cases presented above;
the example with two antisymmetric tensors can be worked out in a
similar fashion.  In each case we introduce Goldstone fields by a
St\"uckelberg redefinition.  We unHiggs the complex massive vector
$\cA_{m}$ by replacing
\eq{
\cA_{m}\ \rightarrow\ \cA_{m} - {\sqrt{2}\over m}\partial_m\phi \ ;
}
for the dual representation, we take
\al{
A_{m} &\rightarrow& A_{m} - {1\over m}\partial_m\phi \nonumber \\
B_{mn}&\rightarrow& B_{mn} - {1\over m}\partial_{[ m} B_{n ]}\ . \nonumber 
}
The introduction of the Goldstino $\nu$ requires an additional
shift
\eq{
\psi_m\ \rightarrow\ \psi_m - {1\over\sqrt{6}m}
(2 \partial_m \nu + \I m \sigma_m \bar\nu) \nonumber 
}
to obtain a proper kinetic term for $\nu$.

In Figure 1(a) the physical fields of the traditional representation
for the massive spin-3/2 multiplet are arranged in terms of 
massless $N=1$ multiplets.  The lowest superspins form an $N=1$ 
chiral and an $N=1$ vector multiplet.  These fields may be thought 
of as $N=1$ ``matter.''  The remaining fields are the gauge fields of
$N=2$ supergravity.   In unitary gauge, the two vectors eat the two
scalars, while the Rarita-Schwinger field eats one linear combination of 
the spin-1/2 fermions.  This leaves the massive $N=1$ multiplet coupled 
to $N=1$ supergravity.  As we shall see, Figure 1 only illustrates 
the field content;  it does not describe the $N=1$ multiplet structure 
of the unHiggsed theory.

\begin{figure}[t]
\epsffile[100 470 530 630]{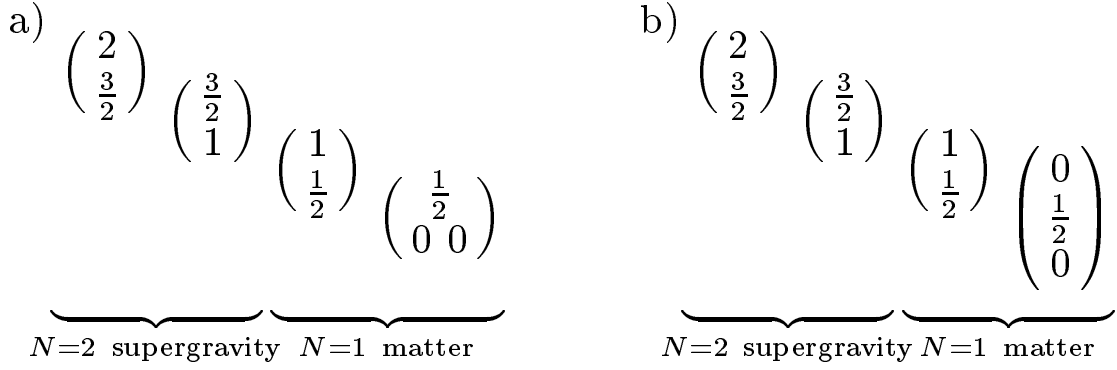}
\caption{The unHiggsed versions of the (a) traditional and (b)
dual representations of the $N=1$ massive spin-3/2 multiplet.}
\end{figure}

The resulting Lagrangian is as follows,
\bea
&& e^{-1}\cL \ =\ \nn\\
&& -\ {1 \over 2 \kappa^2} {\cal R}
 + \epsilon^{m n r s} \overline \psi_{m i}
  \overline \sigma_n D_r \psi^i_s 
 - \I \overline \chi \  \overline \sigma^m D_m \chi 
 - \I \overline \lambda \overline \sigma^m D_m \lambda
-  \cD^m \phi \overline{\cD_m \phi} \nn\\
&& -\ {1 \over 4} \cA_{m n} \overline \cA^{m n} 
 - \ \Bigl( {1 \over \sqrt{2}} m  \psi^2_m \sigma^m
   \overline \lambda 
 + \I m\psi^2_m \sigma^m
   \overline \chi 
+ \sqrt{2}  \I m \lambda \chi 
 +\ {1 \over 2} m \chi \chi \nn\\
&& +\ \Red m \Black\, \psi^2_m \sigma^{m n} \psi^2_n 
+\ {\kappa \over 4} \epsilon_{i j}
\psi^i_m \psi_{n}^{j}
    \overline H_+^{m n} 
 + {\kappa \over \sqrt{2}}  \chi \sigma^m
    \overline \sigma^n \psi^1_m \overline{\cD_n \phi} 
 \nonumber \\
& & + \  {\kappa \over 2 \sqrt{2}}  \overline
\lambda
    \overline \sigma_m \psi^1_n \overline H_-^{m n}
 +  {\kappa \over \sqrt{2}}
    \epsilon^{m n r s} 
    \overline \psi_{m 2} \overline \sigma_n \psi^1_r
    \overline{\cD_s \phi}
    \plushc \Bigr) \ ,
\eea
where $\kappa$ denotes Newton's constant, $m = \kappa v^2$, and $D_m$ is the
covariant derivative.
The supercovariant derivatives take the form
\bea
\hat\cD_m\phi&\ =\ &\partial_m\phi - {\kappa\over\sqrt{2}}\psi^1_m\chi - 
{1\over\sqrt{2}} \kappa v^2  \cA_m\nonumber \\
\hat\cA_{mn}&=&\cA_{mn} + \kappa\psi^2_{ [m}\psi^1_{n] } - 
{\kappa\over\sqrt{2}}\bar\lambda\bar\sigma_{ [n}\psi^1_{m ]} \ .
\eea
This Lagrangian is invariant under two independent abelian gauge 
symmetries, as well as the following supersymmetry transformations,
\newpage
\bea
\delta e^a_m &\ =\ &\I \kappa (\eta^i \sigma^a \overline \psi_{m i} + 
   \bar\eta^i \bar\sigma^a \psi_{m i}) \nonumber \\
\delta \psi^i_{m} & = & {2 \over \kappa} D_m \eta^i
     \nonumber \\
& &   +\ \left( -{\I \over 2} \hat H_{+m n}
\sigma^n \overline \eta^1
 + \sqrt{2} \overline{\cD_m \phi} \eta^1 
 - \kappa\psi^1_m(\bar\chi\bar\eta^1) 
 + \I v^2 \sigma_m \overline \eta^2 \right){\delta_2}^i \nonumber \\
\delta \cA_m &=& 2 \epsilon_{i j} \psi_{m}^{i} \eta^j 
+ \sqrt{2}  \overline \lambda \overline \sigma_m \eta^1 \nonumber \\
\delta \lambda  &=&  {\I \over \sqrt{2}} \overline{\hat\cA}_{mn} 
\sigma^{mn} \eta^1
\Red - \I \sqrt{2} v^2 \eta^2 \Black \nonumber \\
\delta \chi  &=&  \I \sqrt{2} \sigma^m {\hat\cD_m \phi}\overline \eta^1
\Red + 2 v^2 \eta^2 \Black \nonumber \\
\delta \phi &=& \sqrt{2} \chi \eta^1 \ , \label{chiraltrafo}
\eea
for $i=1,2$.  This result holds to leading order, that is, up
to and including terms in the transformations that are linear in
the fields.  Note that this representation is irreducible in the sense 
that there are no subsets of fields that transform only into themselves
under the supersymmetry transformations.

Let us now consider the dual case with one massive tensor. 
The degree of freedom counting is shown in Figure 1(b).  This
time, however, the ``matter'' fields include an $N=1$ vector
multiplet together with an $N=1$ {\it linear} multiplet.  In
unitary gauge, one vector eats one scalar, while the antisymmetric 
tensor eats the other vector.  These are the minimal set of
fields that arise when coupling the alternative spin-3/2 multiplet
to $N=2$ supergravity.

The Lagrangian and 
supersymmetry transformations for this system can be worked out following the
same procedures described above. They
can also be derived by dualizing first the scalar $\phi_B$ and then the vector $B_m$
using the method\footnote{The transformations 
(\ref{vectortrafo}) do not appear to be dual to (\ref{chiraltrafo}),
because the vectors $A_m$ and $B_m$ in (\ref{vectortrafo})
have been rotated to simplify the transformations.} described in \cite{cremdual}. 
As $\kappa
\rightarrow 0$, the
dualities relating a massless antisymmetric tensor $B_{mn}$ to
a massless scalar $\phi$ and a massless vector $A_m$ to another
vector $B_m$ reduce to the simple expressions  $v_m = -\partial_m \phi$ and $F^B_{mn} =
1/2\ \epsilon_{mnrs}F^{Ars}$.

The Lagrangian is as follows,
\bea
& &e^{-1}\cL\ =\ \nonumber \\
&  &  -\ {1 \over 2 \kappa^2} {\cal R}
 + \epsilon^{pqrs} \bar \psi_{p i}
  \bar \sigma_q D_r \psi^i_s 
 - \I \bar \chi \bar \sigma^m D_m \chi 
 - \I \bar \lambda \bar \sigma^m D_m \lambda 
-  {1\over 2}\cD^m \phi \cD_m \phi \nn\\
&& -\ {1 \over 4} \cF^A_{mn} \cF^{Amn} - {1\over4}\cF^B_{mn}\cF^{Bmn}
+ {1\over 2}v^m v_m
  - \Bigl( {1 \over \sqrt{2}}  m \,  \psi^2_m \sigma^m
   \bar \lambda  + m \I \psi^2_m \sigma^m
   \bar \chi 
 \nonumber \\
& &+\ \sqrt{2} m \I \lambda \chi 
 + {1 \over 2} m \chi \chi 
 + \Red m \Black\, \psi^2_m \sigma^{m n} \psi^2_n  
+ {\kappa \over 2\sqrt{2}}  \epsilon_{i j} \psi^i_m \psi_{n}^{j}
    \cF^{Amn}_{-}   \nn\\
    &&+\ {\kappa \over {2}}  \chi \sigma^m
    \bar \sigma^n \psi^1_m \cD_n \phi
+  {\kappa \over 2 }  \bar \lambda
    \bar \sigma_m \psi^1_n \cF^{Bmn}_{+}
    +
  {\kappa \over {2}} 
    \epsilon^{pqrs}
    \bar \psi^2_{p } \bar \sigma_q \psi^1_r
    \cD_s \phi\nn\\
&&- \I\, {\kappa \over {2}}  \chi \sigma^m
    \bar \sigma^n \psi^1_m v_n 
-  \I \, {\kappa \over {2}} 
    \epsilon^{pqrs}
    \bar \psi^2_{p } \bar \sigma_q \psi^1_r
    v_s
    \plushc \Bigr)
\eea
where, as before, $m = \kappa v^2$, and
\bea
\cD_m \phi &\ =\ & \partial_m \phi - {m\over\sqrt{2}} (A_m + B_m) \nonumber \\
\cF^A_{mn} &=& \partial_{[m }A_{n]} + {m\over\sqrt{2}} B_{mn} \nonumber \\
\cF^B_{mn} &=& \partial_{[m }B_{n]} - {m\over\sqrt{2}} B_{mn}\ . 
\eea
This Lagrangian is invariant under an ordinary abelian gauge symmetry,
an antisymmetric tensor gauge symmetry, as well as the following two 
supersymmetries,
\bea
\delta_{\eta} e^a_m &\ =\ &\I\, \kappa (\eta^i \sigma^a \overline \psi_{m i} + 
    \bar\eta^i \bar\sigma^a \psi_{m i}) \nonumber \\
\delta_{\eta} \psi^1_m &=& {2\over\kappa}D_m\eta^1 \nonumber \\
\delta_\eta A_m &=& \sqrt{2}\epsilon_{ij}(\psi_m^i\eta^j + \bar\psi_m^i\bar\eta^j) \nonumber \\
\delta_\eta B_m &=& \bar\eta^1\bar\sigma_m\lambda + \bar\lambda\bar\sigma_m\eta^1\nonumber \\
\delta_\eta B_{mn} &=& 2\eta^1\sigma_{mn}\chi 
+ \I\,\eta^1\sigma_{[ m}\bar\psi^2_{n ]}
+ \I\,\eta^2\sigma_{[ m}\bar\psi^1_{n ]}
\plushc \nonumber\\
\delta_\eta \lambda &=& \I\, \hat\cF^B_{mn}\sigma^{mn}\eta^1 - \I \sqrt{2}
v^2 \eta^2\nonumber\\
\delta_\eta \chi &=& \I\, \sigma^m\bar\eta^1 \hat\cD_m\phi - \hat
v_m\sigma^m\bar\eta^1 + 2 v^2 \eta^2\nonumber \\
\delta_{\eta} \psi^2_m &=& {2\over\kappa}
D_m\eta^2 + \I v^2 \sigma_m \bar\eta^2 
- {\I\over\sqrt{2}}\hat\cF^A_{+mn}\sigma^n\bar\eta^1\nonumber \\
   & & +\ \hat\cD_m \phi \eta^1 +\kappa\left( (\bar\psi^1_m\bar\chi)\eta^1
- (\bar\chi\bar\eta^1)\psi^1_m \right) - \I\, \hat v_m\eta^1\nonumber \\
\delta_\eta \phi &=& \chi\eta^1 + \bar\chi\bar\eta^1 \label{vectortrafo}
\eea
up to linear order in the fields.
The supercovariant derivatives are given by
\bea
\hat \cD_m\phi&\ =\ & \cD_m\phi - {\kappa\over{2}}(\psi^1_m\chi +
\bar\psi^1_m\bar\chi)  \nonumber\\
\hat \cF^A_{mn}&=& \cF^A_{mn} + {\kappa\over \sqrt{2}}(\psi^2_{
[m}\psi^1_{n] } + \bar\psi^2_{ [m}\bar\psi^1_{n] }) \nonumber \\
\hat \cF^B_{mn}&=& \cF^B_{mn} - {\kappa\over 2}(\bar\lambda\bar\sigma_{
[n}\psi^1_{m ]} + \bar\psi^1_{ [m}\bar\sigma_{n] }\lambda)  \nonumber\\
\hat v_m &=& v_m + \Bigl(\, \I\kappa \psi^{1}_n\sigma_{m}{}^n\chi  
                 -{\I\kappa\over
2}\epsilon_{m}{}^{nrs}\psi_n^{1}\sigma_r\bar\psi_s^{2} 
\plushc \Bigr)  \ .
\eea
These fields form an irreducible representation of the $N=2$
algebra.

Each of the two Lagrangians has a full $N=2$ supersymmetry (up to
the appropriate order).  The first
supersymmetry is realized linearly.  The second
is realized nonlinearly: it is spontaneously broken.  In each case,
the transformations imply that
\be
\zeta\ =\ {1\over \sqrt3}\,
(\chi - \I \sqrt 2 \lambda)
\ee
does not shift, while
\be
\nu\ =\ {1\over \sqrt3}\,
(\sqrt 2 \chi + \I \lambda )
\ee
does. Therefore $\nu$ is the Goldstone fermion for $N=2$ supersymmetry,
spontaneously broken to $N=1$.

\section{Dual Algebras from Partial Supersymmetry Breaking}

Now that we have explicit realizations of partial supersymmetry
breaking, we can see how they avoid the no-go argument
presented in the introduction.  We first compute the second
supercurrent.  In each case it turns out to be
\be
J^2_{m\alpha}\ =\ 
v^2\,(\sqrt6\, \I \,\sigma_{\alpha\dot\alpha m}
\bar\nu^{\dot\alpha} +
4\,\sigma_{\alpha\beta m n} \psi^{2n\beta})\ ,
\ee
plus higher-order terms.  The commutator of the second
supercharge with the second supercurrent is then
\be
\{\,\bar S_{\dot\alpha},\,J^2_{m\alpha}
\,\}\ =\ 0 \ +\ \hbox{terms at least linear in the fields}\ .
\ee
From this we see that the stress-energy tensors in the current
algebra (\ref{zweite}) do not differ by a constant shift.
The supergravity couplings must exploit the second loophole
to the no-go theorem.

To check this assertion, note that the 
operators $J^i_{\alpha m}$ and $T_{mn}$ contain contributions
from {\it all} of the fields, including the second gravitino.  When
covariantly-quantized, the second gravitino gives rise to states of
negative norm. Indeed, we find
\be
(\bar S S + S \bar S)\,
|0\rangle\ =\ 0\ ,
\ee
even though
\be
S\,|0\rangle\ \ne\ 0 \quad\qquad
\bar S\,|0\rangle\ \ne\ 0\ .
\ee

To elucidate the role of the bosonic symmetries associated with
partial supersymmetry breaking, let us now compute the closure of
the first and second supersymmetry transformations to zeroth order
in the fields.  In this way we can identify the Goldstone fields
associated with any spontaneously broken bosonic symmetries.

For the traditional representation, (Figure 1(a)), we find
\bea
\left[ \,\delta_{\eta_1}, \,\delta_{\eta_2} \right] \, \phi 
&\ =\ &  2\sqrt{2}\,  v^2\,\eta_1\eta_2  \nonumber \\
\left[ \,\delta_{\eta_1}, \,\delta_{\eta_2} \right] \, {\cal A}_m
&=& {4\over\kappa} \, \partial_m\, (\eta_1\eta_2)\ .
\eea
This shows that the complex scalar $\phi$ is indeed the Goldstone
boson for a gauged central charge.  Moreover, in unitary gauge,
where
\be
\phi\ =\ \nu\ =\ 0\ ,
\ee
this Lagrangian reduces to the usual representation for a massive
$N=1$ spin-3/2 multiplet \cite{fvn}.

For the dual representation (Figure 1(b)), we have
\bea
\left[ \,\delta_{\eta_1}, \,\delta_{\eta_2} \right] \, \phi 
&\ =\ &  2\,  v^2\,(\eta_1\eta_2 + \bar\eta_1\bar\eta_2)  \nonumber \\
\left[ \, \delta_{\eta^2}, \,  \delta_{\eta^1} \right] \,  A_m &\ =\ &
{2\sqrt{2} \over \kappa}
\partial_m(\eta^1\eta^2 + \bar\eta_1\bar\eta_2) 
  - \sqrt{2} \, \I \,  v^2 \, (\eta^2\sigma_m\bar\eta^1 -
\eta^1\sigma_m\bar\eta^2) \nonumber \\
\left[ \,  \delta_{\eta^2}, \,  \delta_{\eta^1} \right] \,  B_m &=& 
\sqrt{2} \, \I \,  v^2 \, (\eta^2\sigma_m\bar\eta^1 -
\eta^1\sigma_m\bar\eta^2) \nonumber \\
\left[ \,  \delta_{\eta^2}, \,  \delta_{\eta^1} \right] \,  B_{mn} &=&
{2 \, \I\over \kappa}D_{[m }
(\eta^2\sigma_{ n]}\bar\eta^1 - \eta^1\sigma_{n] }\bar\eta^2)\ .
\eea
The real vector $-(A_m - B_m)/\sqrt{2}$ is the Goldstone boson for
a gauged {\it vectorial} central extension of the $N=2$ algebra.  In
addition, the real scalar $\phi$ is the Goldstone boson associated
with a single real gauged central charge.  In unitary gauge, with
\be
-{1\over\sqrt{2}}(A_m - B_m)\ =\ \phi\ =\ \nu\ =\ 0\ ,
\ee
this Lagrangian reduces to the dual representation for
the massive $N=1$ spin-3/2 multiplet \cite{ogsok}.

Finally, for the case with two tensors $\cA_{mn}= A_{mn} +\I B_{mn}$
and two Goldstone vectors $\cA_m = A_m +\I B_m$, the algebra is
\al{
\left[ \delta_{\eta^2}, \delta_{\eta^1} \right] \cA_m\ &=&\ {4\over \kappa}
D_m(\bar\eta^1\bar\eta^2) - 4\I v^2 \eta^2\sigma_m\bar\eta^1 \nonumber \\
\left[ \delta_{\eta^2}, \delta_{\eta^1} \right] \cA_{mn} &=& -{4\I\over 
\kappa}D_{[m }
(\eta^2\sigma_{ n]}\bar\eta^1), \nonumber
}
This case requires {\it two} vectorial central extensions
of the supersymmetry algebra.

\section{Discussion and Conclusion}

In this paper we have examined the partial breaking of supersymmetry
in flat space.  We have seen that partial breaking can be accomplished
using either of three representations of the massive $N=1$ spin-3/2
multiplet.  We unHiggsed the representations, and found
a new $N=2$ supergravity and a new $N=2$ supersymmetry algebra.  

Each of these theories gives rise to different $N=1$ multiplet
structures in the limit $\kappa \rightarrow 0$.  For the
traditional representation, we find a massless chiral multiplet,
($\chi$, $\phi$), together with a pair of ``twisted" massless $N=1$
multiplets, ($\psi^2_m$, $\cA_m$, $\lambda$).  The twisted multiplets
transform irreducibly into each other under the first, unbroken 
supersymmetry.  They can be untwisted with the help of a second
unbroken supersymmetry which appears in this limit.\footnote{We
are indebted to W.~Siegel for pointing this out.}  The second
supersymmetry transformations are obtained from (\ref{chiraltrafo})
(in the $\kappa \rightarrow 0$ limit) by $\cA_m \rightarrow \bar\cA_m$,
$\lambda \rightarrow -\lambda$.  We see that the twisted multiplet
is actually a massless $N=2$ multiplet.

In the case of the dual representation, the $N=1$ transformations
(\ref{vectortrafo}) reduce, in the $\kappa \rightarrow 0$ limit,
to those of a massless vector multiplet, ($B_m$, $\lambda$), a
linear multiplet, ($\chi$,  $B_{mn}$, $\phi$), and a massless
spin-3/2 multiplet, ($\psi^2_m$, $A_m$).\footnote{The transformations
that mix the gravitino and the
antisymmetric tensor are physically irrelevant because the
transformations of the corresponding field strengths vanish
on-shell.}
This multiplet structure can also be obtained by an explicit
superfield unHiggsing of the massive spinor superfield $\Psi_\alpha$
in the Ogievetsky/Sokatchev formulation \cite{ogsok},
\eq{
\cL \ =\  \underbrace{-{1\over 2}(\Psi \bar\Psi) \pi^\perp \pmatrix{ \Psi \cr \bar\Psi \cr}}_{\cL^\perp} +  
{1\over 2}m(\Psi\Psi + \bar\Psi\bar\Psi) \ , \label{perpm}
}
where $\pi^\perp = \sqrt{\Box \Pi_1}$ and $\Pi_1$ is the superspin-1
projector for a spinor superfield \cite{sok}. 
The St\"uckelberg redefinition $ \Psi_\alpha \rightarrow \Psi_\alpha
- \I D_\alpha V + L_\alpha + 2\I W_\alpha/m + D_\alpha L/4m$ leads to
\eq{
\cL\ \rightarrow\ \cL^\perp + {2}W^\alpha D_\alpha V - {1\over 8}L^2 + {1\over 2}m(
(\Psi_\alpha - \I D_\alpha V + L_\alpha)^2 \plushc)\ , \nonumber
}
where $V$ is a real vector and $L_\alpha$ a chiral spinor superfield; $W_\alpha$ and $L$ are the corresponding field strengths.  The correct
multiplet structure is obtained in the limit $m \rightarrow 0$.
Note, however, that the auxiliary superspin-0 is lost so we expect
${1/m}$ singularities in the supersymmetry transformations. 

The multiplet structure of the dual theory with two antisymmetric
tensors consists of the $N=2$ representation, ($\psi^2_m$, $A_m +\I B_m =
\cA_m$, $\lambda$), as well as a linear multiplet with two antisymmetric
tensors, ($\chi$, $A_{mn} +\I B_{mn} = \cA_{mn}$). 
The argument that prevents the coupling of this multiplet to
supergravity (see \cite{nonclos} and references therein) does not apply here
since the ``non-closure'' terms in the supersymmetry algebra are cancelled by
terms from the variation of $\psi^2_m$.

The Lagrangian for the traditional representation is a truncation of
the supergravity coupling found by Cecotti, Girardello, and Porrati,
and by Zinov'ev \cite{zin}.
Their results were based on {\it linear} $N=2$ supersymmetry; they
involved at least one $N=2$ vector-multiplet and one hypermultiplet.
The Lagrangians for the dual cases are new. They contain new realizations
of $N=2$ supergravity.

In each case, the couplings presented here are minimal and model-independent.
They describe the superHiggs effect in the low-energy effective theories
that arise from partial supersymmetry breaking.

\section{Acknowledgements}
We would like to thank S.~Osofsky for collaboration at an early stage of 
this project and A.~Galperin and W.~Siegel for useful discussions.  
This work was supported by the National Science
Foundation, grant NSF-PHY-9404057.

\end{document}